\documentclass[%
 reprint,
 amsmath,amssymb,
 aps,
 prl,
superscriptaddress,
 floatfix,
]{revtex4-2}

\usepackage{graphicx}
\usepackage{dcolumn}
\usepackage{bm}
\usepackage{multirow}
\usepackage{array}
\usepackage{color}
\usepackage[hidelinks,linktoc=all]{hyperref}

\begin{document}

\title{Dirac Cones in $d$-wave Altermagnets Enable High-Conductivity and High-Efficiency Spin Sources}

\author{Tianye Yu}
\thanks{These authors contributed equally to this work.}
\affiliation{Shenyang National Laboratory for Materials Science, Institute of Metal Research, Chinese Academy of Sciences, Shenyang 110016, China.}

\author{Junwen Lai}
\thanks{These authors contributed equally to this work.}
\affiliation{Shenyang National Laboratory for Materials Science, Institute of Metal Research, Chinese Academy of Sciences, Shenyang 110016, China.}
\affiliation{School of Materials Science and Engineering, University of Science and Technology of China, Shenyang 110016, China.}

\author{Peitao Liu}
\affiliation{Shenyang National Laboratory for Materials Science, Institute of Metal Research, Chinese Academy of Sciences, Shenyang 110016, China.}

\author{Xing-Qiu Chen}
\email{xingqiu.chen@imr.ac.cn}
\affiliation{Shenyang National Laboratory for Materials Science, Institute of Metal Research, Chinese Academy of Sciences, Shenyang 110016, China.}

\author{Yan Sun}
\email{sunyan@imr.ac.cn}
\affiliation{Shenyang National Laboratory for Materials Science, Institute of Metal Research, Chinese Academy of Sciences, Shenyang 110016, China.}

\date{\today}



\begin{abstract}
Low critical charge-current density and low energy dissipation are highly desired in magnetic random-access memories, requiring spin sources to exhibit both high charge-to-spin conversion efficiency (CSE) and high charge conductivity. Altermagnets with vanishing net magnetic moment and spin-splitting bands provide promising spin-source candidates for spin-splitting-torque magnetic random-access memories. However, achieving both high CSE and charge conductivity remains challenging in altermagnets. In this work, we introduce Dirac cones into two-dimensional $d$-wave altermagnets, where their intrinsically high carrier mobility enables tunable charge and spin conductivities with high CSE. Dirac-cone anisotropy provides an effective means of enhancing both CSE and charge conductivity, with cone tilting serving as an additional degree of tunability. Guided by this design principle, we identify a maximum CSE of 92\% in Cr$_2$SeTeS. When the Fermi level moves slightly away from the Dirac point, high CSE, high charge conductivity, and the resulting high spin conductivity can be simultaneously achieved. Our study advances the understanding of time-reversal-odd spin transport via Dirac-cone engineering and provides a practical route toward developing spin-source materials that combine high charge conductivity with highly efficient charge-to-spin conversion.

\end{abstract}

\pacs{Valid PACS appear here}
\maketitle
\textit{Introduction}---Current-induced spin torques provide an efficient route for writing magnetic bits in spintronic memories \cite{miron2011perpendicular,liu2012tantalum,apalkov2016sttmram,manchon2019sot,guo2021spintronics,yang2022twod,fert2024electrical}. The key component is the spin source, which converts an applied charge current into a spin current. Ferromagnetic spin sources can deliver large spin polarization, but their net magnetization also brings stray fields, magnetic crosstalk, and limited scalability \cite{wadley2016electrical,jungwirth2016afspintronics,baltz2018afspintronics,godinho2024domainwall}. This has motivated the search for spin sources with vanishing net magnetic moment.

\begin{figure}
\includegraphics[width=3.5in]{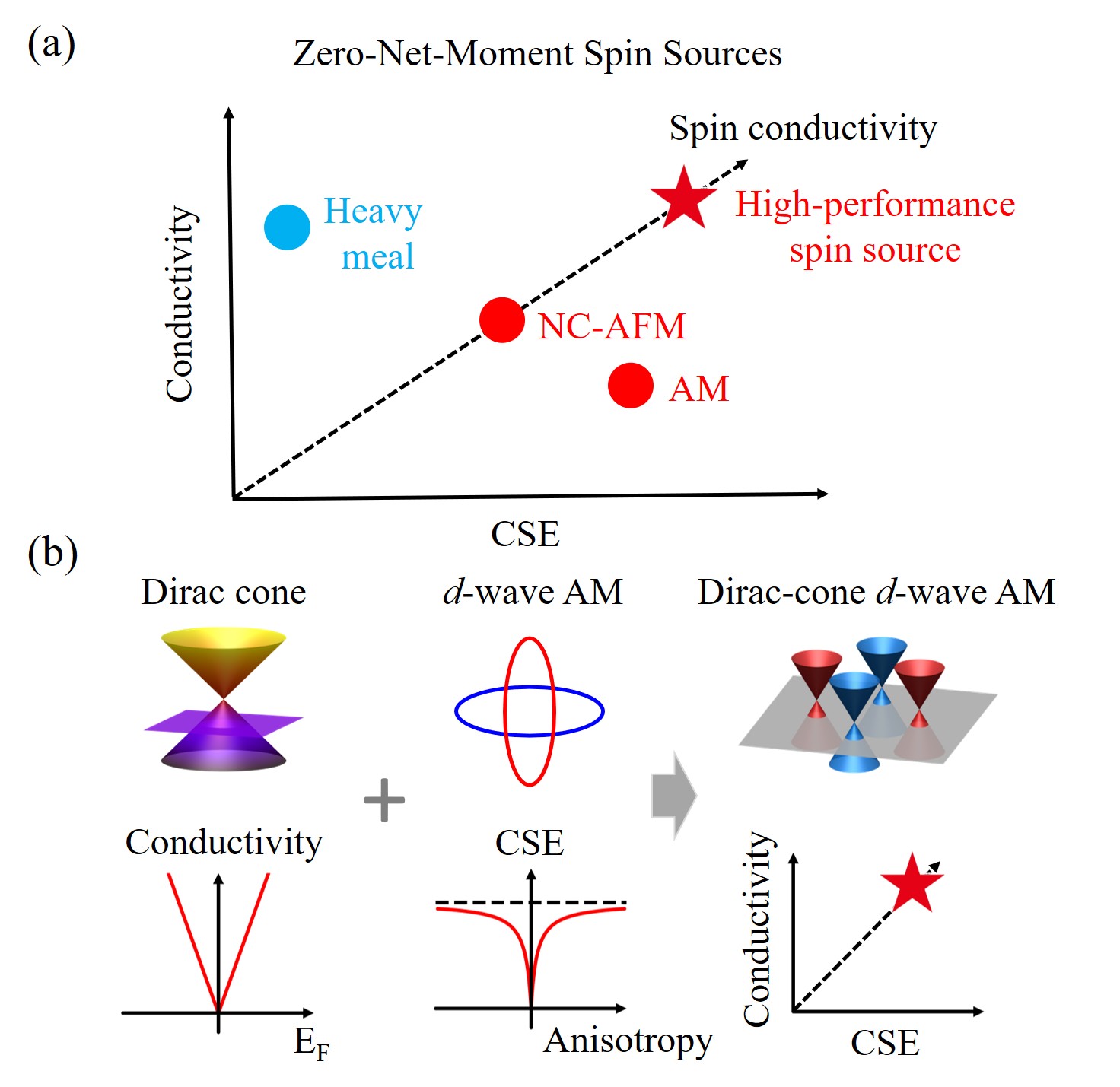}
\caption{\label{fig:1}
Schematic comparison of representative spin-source materials with vanishing net magnetic moment and the proposed route toward a high-performance spin source. (a) Comparison of the charge conductivity and charge-to-spin conversion efficiency (CSE) of heavy metals, noncollinear antiferromagnets (NC-AFMs), altermagnets (AMs), and ideal spin sources. The diagonal direction represents increasing spin conductivity, since the spin conductivity is given by the product of the charge conductivity and the CSE. Blue and red denote spin sources dominated by the $T$-even spin Hall current arising from spin-orbit coupling and the $T$-odd spin current arising from spin splitting, respectively. (b) Combining the high charge conductivity enabled by the intrinsically high carrier mobility of Dirac cones with the high CSE attainable in $d$-wave AMs provides a route toward a high-performance spin source. The CSE can be further enhanced by increasing the anisotropy of the Dirac cones.
}
\end{figure}

Heavy metals, noncollinear antiferromagnets, and altermagnets have emerged as representative zero-net-moment candidates \cite{sinova2015spinhalleffects,ma2021multifunctional,shao2021roadmap,hu2022magneticsh,smejkal2022altermagnetism,smejkal2022emerging,bai2024altermagnetism,song2025altermagnets,jungwirth2026symmetry}. Their performance is governed by the spin conductivity $\sigma^s$, which can be written as $\sigma^s=\mathrm{CSE}\times\sigma$, where $\sigma$ is the charge conductivity and CSE is the charge-to-spin conversion efficiency \cite{khang2018conductive,li2019materials,zhu2020energy,xu2020ptte2,hibino2021giant,hu2022magneticsh,wang2023fegteti}. A large $\sigma$ with a small CSE still requires a high critical current to generate sufficient torque \cite{ramaswamy2018recent}, whereas a large CSE with a small $\sigma$ produces voltage drop, Joule heating, and resistance-capacitance delay \cite{zhu2020energy}. Thus, an ideal spin source should simultaneously exhibit high charge conductivity and a high CSE \cite{khang2018conductive,zhu2018aupt,zhu2019pdpt}. However, few zero-net-moment candidates can meet both requirements. Specifically, heavy metals are highly conductive but typically exhibit limited CSE \cite{liu2012tantalum,sinova2015spinhalleffects,zhu2020energy,shao2021roadmap,krizakova2022sotmram,nguyen2024sotmram}. Altermagnets can generate a high CSE through their spin-splitting bands, but they often suffer from low charge conductivity, with many reported altermagnets being insulating or semiconducting \cite{guo2023spinsplit,gao2025ai,wan2025htsearch,sufyan2026quantification,guo2026monolayers}. Noncollinear antiferromagnets, by contrast, generally exhibit moderate charge-to-spin conversion efficiency and charge conductivity \cite{zelezny2017spinpolarized,hu2022magneticsh}, as summarized in Fig.~\ref{fig:1}(a). There is therefore an urgent need to establish an effective strategy for realizing high-performance spin-source materials.

The central challenge is to make an altermagnetic spin source highly conductive without sacrificing its spin-conversion efficiency. Since $\sigma=ne\mu$, increasing the carrier density $n$ can improve conductivity, but excessive doping may introduce ordinary metallic carriers or move the Fermi level away from the band region responsible for efficient spin conversion. A cleaner route is to enhance the carrier mobility $\mu$. Dirac cones are well suited for this purpose because their linear dispersion, high Fermi velocity, and small effective mass naturally favor high mobility \cite{burkov2016topological,armitage2018Dirac,lv2021Dirac}. When the Fermi level is shifted slightly away from the Dirac point but remains within the linear-dispersion regime, high conductivity can be obtained while preserving the local band character, as exemplified by graphene, whose conductivity can be readily tuned through gate voltage \cite{novoselov2004electric,zhang2005experimental,das2008monitoring}. Altermagnets provide the complementary ingredient, their ferromagnet-like spin-polarized bands with zero net moment can transfer charge current into time-reversal-odd ($T$-odd) spin current without relying on spin-orbit coupling. Instead, their spin-splitting bands can give rise to high CSE \cite{gonzalez2021efficient,smejkal2022altermagnetism,lai2025dwave}. Embedding Dirac cones into $d$-wave altermagnets therefore offers a route to spin sources with both high charge conductivity and high CSE, as shown in Fig.~\ref{fig:1}(b).

In this work, we establish Dirac-cone engineering as an effective strategy for optimizing altermagnetic spin sources. Model analysis shows that Dirac-cone anisotropy strongly enhances the CSE, while type-I cone tilting provides an additional tuning knob. Guided by this principle, we examine Cr$_2$O$_2$ and Cr$_2$SeTeS, two theoretically proposed 2D $d$-wave altermagnets \cite{guo2023qah,chen2023cro,xu2026monolayer}. Cr$_2$O$_2$ realizes a weakly tilted anisotropic Dirac cone and exhibits a CSE of 72\%. Cr$_2$SeTeS combines stronger anisotropy with sizable tilt, yielding a CSE of 92\% at the charge-neutrality point. Upon shifting the Fermi level upward by only 26 meV, Cr$_2$SeTeS retains a CSE as high as 94\% while reaching a $T$-odd spin conductivity on the order of $10^3(\hbar/e)(\mathrm{S/cm})$. These results identify Dirac-cone engineering in $d$-wave altermagnets as a practical route toward spin sources that are simultaneously conductive and highly efficient.

\begin{figure*}
\includegraphics[width=7.0in]{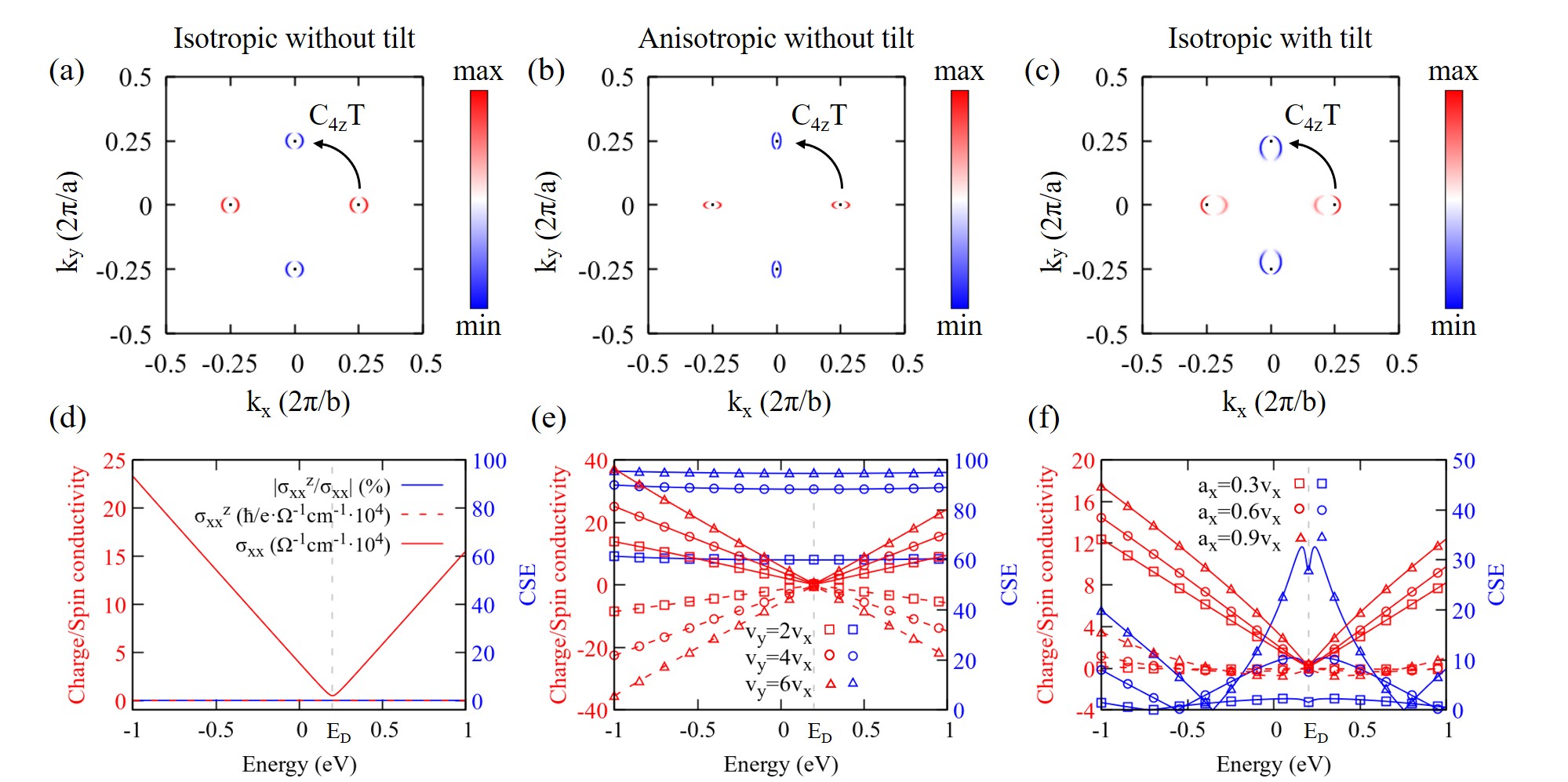}
\caption{\label{fig:model_am} Model-calculated $k$-resolved $T$-odd spin-current distribution and Fermi-level-dependent electrical conductivity, spin conductivity, and CSE in a two-dimensional altermagnet (AM). (a) Isotropic Dirac cone in a 2D AM, (b) anisotropic Dirac cone in a 2D AM, and (c) isotropic type-I Dirac cone with tilt. Black points denote the Dirac points. (d)--(f) The corresponding electronic conductivity, spin conductivity, and CSE obtained with different amplitudes of anisotropy and tilt in (a)--(c). The gray dashed line denotes the Dirac-point energy.}
\end{figure*} 

\begin{figure*}
\includegraphics[width=7in]{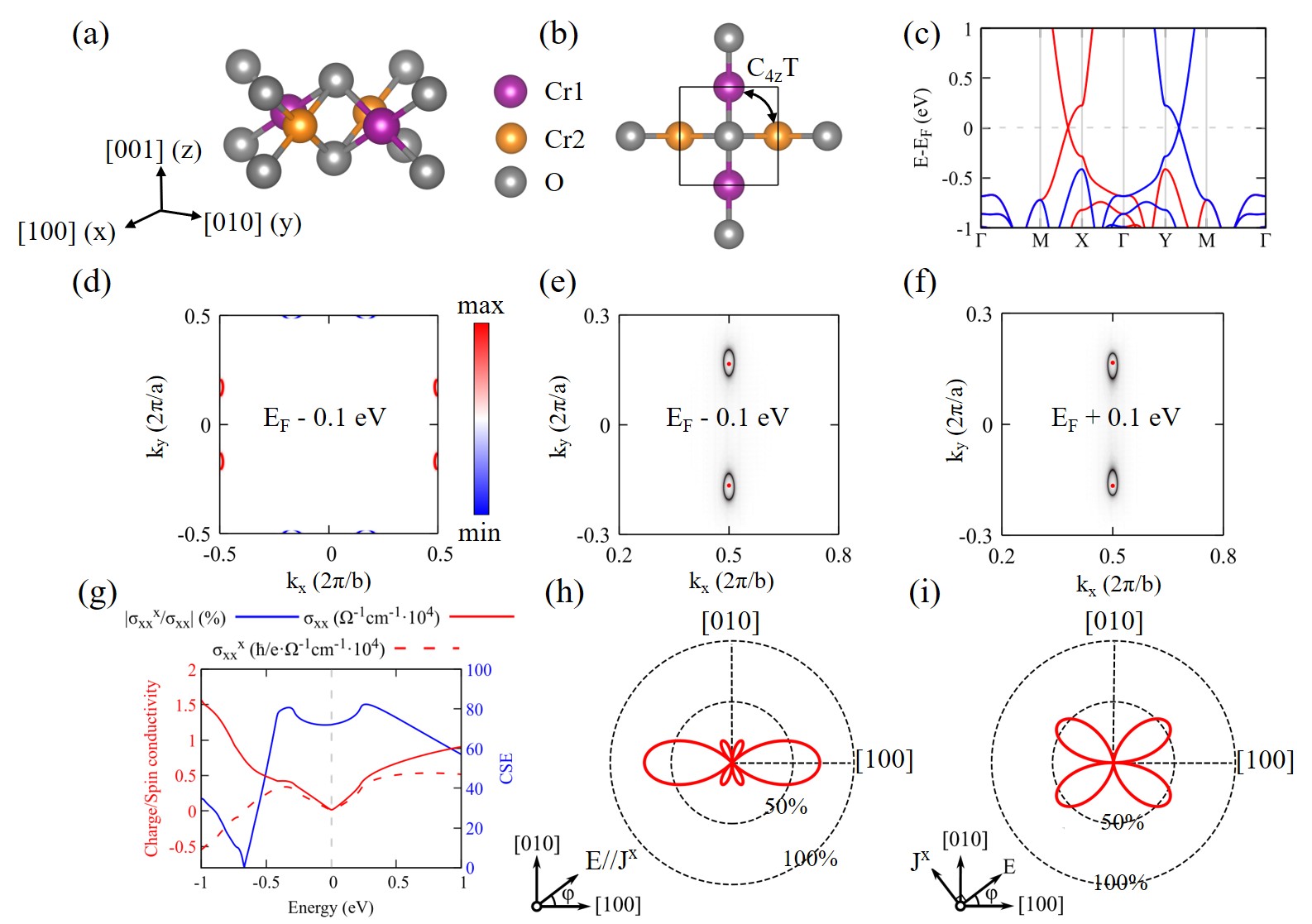}
\caption{\label{fig:cr2o2} Crystal structure, electronic structure, and spin transport properties of Cr$_2$O$_2$. Crystal structure of Cr$_2$O$_2$ viewed from (a) the side view and (b) the top view. Cr1 and Cr2 label Cr atoms with oppositely oriented magnetic moments. (c) Spin-resolved band structure without spin-orbit coupling (SOC), red and blue lines indicate different spin orientations. (d) $k$-resolved $T$-odd spin current. (e),(f) Fermi surfaces with the Fermi level shifted downward and upward by 0.1 eV, red points denote Dirac points. (g) Fermi-energy-dependent electronic conductivity, spin conductivity, and CSE. (h),(i) Angular dependence of the longitudinal and transverse CSE at the charge neutrality point.}
\end{figure*} 

\textit{Results and discussion}---Under an applied electric field, the nonequilibrium redistribution of spin-splitting Fermi surfaces produces an intraband spin-polarized current. The spin conductivity tensor is evaluated as \cite{freimuth2014spinorbit,gonzalez2021efficient}
\begin{equation}
\sigma_{ij}^{k}=-\frac{e\hbar}{V\pi N_k}\sum_{\mathbf{k}}\tilde{\sigma}_{ij}^{k}(\mathbf{k}),
\end{equation}
\begin{equation}
\tilde{\sigma}_{ij}^{k}(\mathbf{k})=
\sum_{mn}
\frac{
\mathrm{Re}\left[
\langle u_{n\mathbf{k}}|\hat{J}_{i}^{k}|u_{m\mathbf{k}}\rangle
\langle u_{m\mathbf{k}}|\hat{v}_{j}|u_{n\mathbf{k}}\rangle
\right]\eta^2
}
{\left[(E_F-E_{n\mathbf{k}})^2+\eta^2\right]
\left[(E_F-E_{m\mathbf{k}})^2+\eta^2\right]},
\end{equation}
where $\hat{J}_{i}^{k}=\{\hat{s}_k,\hat{v}_i\}/2$ is the spin-current operator with spin polarization along $k$ and flow direction along $i$, $\hat{v}_j$ is the velocity operator along the electric-field direction, and $\eta$ is a broadening parameter associated with the relaxation time. The charge-to-spin conversion efficiency is defined as $\mathrm{CSE}=\sigma_{ij}^{k}/\sigma_{ii}\times100\%$, where $\sigma_{ii}$ is the charge conductivity along the applied field. Based on the above equations, we calculated the spin-transport properties of the Dirac-cone $d$-wave altermagnets.

We first explain the convention for single isolated Dirac cone that we used before embedding the cones into a $d$-wave altermagnetic symmetry setting. The universal 2D $k\cdot p$ Hamiltonian can be written as \cite{burkov2016topological,armitage2018Dirac,lv2021Dirac}
\begin{equation}
    H=\sum_{ij}v_{ij}k_i\sigma_j+(a_ik_i-\mu)\sigma_0,
\end{equation}
where $\sigma_j$ are Pauli matrices, $v_{ij}$ describes the Dirac velocity matrix, $a_i$ denotes the tilt, and $\mu$ is the chemical potential. To make the dependence on anisotropy and tilt transparent, we take a minimal form in which the off-diagonal velocity components are omitted and the tilt is constrained along $k_x$. The final lattice-regularized Hamiltonian used in the calculation is
\begin{equation}
\begin{aligned}
H &= v_x \sin(k_x)\sigma_x 
+ v_y \sin(k_y)\sigma_y 
+ [a_x \sin(k_x)-\mu]\sigma_0  \\
&\quad + [2-\cos(k_x)-\cos(k_y)]\sigma_z,
\end{aligned}
\end{equation}
where the ratio between $v_x$ and $v_y$ measures the anisotropy and $a_x$ measures the tilt along $k_x$. The replacement $k_i\rightarrow\sin (k_i)$ has been used to project the continuum model onto a square lattice for Brillouin-zone integration, with the lattice constant chosen to be 1 $\text{\AA}$. Because this regularization also generates extra Fermi surfaces due to the fermion doubling problem (Fig.~S2 in Supplemental Material~\cite{supplemental}), which is not an intrinsic feature of the target $k\cdot p$ Dirac cone, the Wilson mass term $[2-\cos(k_x)-\cos(k_y)]\sigma_z$ is introduced to remove the additional low-energy Fermi surface \cite{stacey1982eliminating,zhou2017surface,resende2017confinement,beenakker2023tangent}.

Our quantitative analysis is restricted to Dirac cones in the type-I regime. For a type-II cone, the electron- and hole-like Fermi pockets extend well beyond the vicinity of the Dirac point, where the local $k \cdot p$ Hamiltonian no longer provides an accurate description of the full band dispersion (Fig. S1 in Supplemental Material ~\cite{supplemental}). Consequently, the Fermi-surface-induced $T$-odd spin current cannot be reliably evaluated within this low-energy model.

We next embed the Dirac cones into a 2D $d$-wave altermagnetic configuration. In conventional collinear antiferromagnets, the opposite-spin Fermi surfaces are degenerate at each $k$ point, and the $T$-odd spin current cancels. In a $d$-wave altermagnet, however, opposite spin sectors are related by the combined $[C_2||C_{4z}]$ operation, where $C_2$ acts on the spin moment and $C_{4z}$ rotates the lattice by $90^\circ$ about the $z$ axis. We therefore place four symmetry-related Dirac cones in the square Brillouin zone, with the spin direction assumed to be along $z$ or $-z$. For each pair of Fermi-surface sectors connected by $C_{4z}$, the net response is determined by the sum $\tilde{\sigma}_{ii}^{k}(k)+\tilde{\sigma}_{ii}^{k}(C_{4z}k)$. The absolute position of the Dirac point is not essential as long as all four cones remain connected by the $C_{4z}$ operation. For concreteness, we place the four Dirac cones at $(\pm 0.25,0)$ and $(0,\pm 0.25)$ in the model calculations.

As shown in Fig.~\ref{fig:model_am}(a), when the Dirac cones are isotropic and untilted, the $C_{4z}$-related sectors have identical circular Fermi-surface geometry, and their opposite-spin contributions exactly cancel, namely $\tilde{\sigma}_{ii}^{k}(k)+\tilde{\sigma}_{ii}^{k}(C_{4z}k)=0$. Consequently, the spin conductivity and CSE remain zero even though the charge conductivity increases as the chemical potential moves away from the Dirac point [Fig.~\ref{fig:model_am}(d)]. A finite $T$-odd spin current therefore requires a Dirac-cone deformation that makes the two symmetry-related Fermi-surface contributions uncompensated.

Anisotropy provides the most direct way to create this imbalance. Once the circular Fermi surface is deformed into an ellipse, the $C_{4z}$-related sectors no longer cancel completely, giving a finite $\sigma_{xx}^{z}$ [Fig.~\ref{fig:model_am}(b)]. As the Fermi level moves away from the Dirac point, both charge and spin conductivities increase monotonically, while the CSE remains constant [Fig.~\ref{fig:model_am}(e)]. Increasing the anisotropy further enhances the CSE. For example, the CSE is about 60\% when $v_y=2v_x$ and approaches the 100\% limit when $v_y=6v_x$. Thus, anisotropy is an efficient route to simultaneously enhance charge conductivity, CSE, and spin conductivity.

Tilt offers an additional but more delicate tuning channel. For an isotropic type-I cone with finite tilt, the displaced Fermi surface also breaks the exact cancellation between $C_{4z}$-related sectors and generates a nonzero spin current [Fig.~\ref{fig:model_am}(c)]. In this case, moving the Fermi level away from the Dirac point first increases the spin conductivity and CSE, but the response subsequently decreases as the chemical potential moves further away from the local Dirac-cone regime [Fig.~\ref{fig:model_am}(f)]. Increasing $a_x$ enhances the charge conductivity, spin conductivity, and CSE within the relevant low-energy window, although the effect is weaker than that of anisotropy. The CSE is only about 30\% for $a_x=0.9v_x$. These results indicate that, for material design, anisotropy should be the primary target, with tilt serving as an additional means to tune the response.

\begin{figure*}
\includegraphics[width=7in]{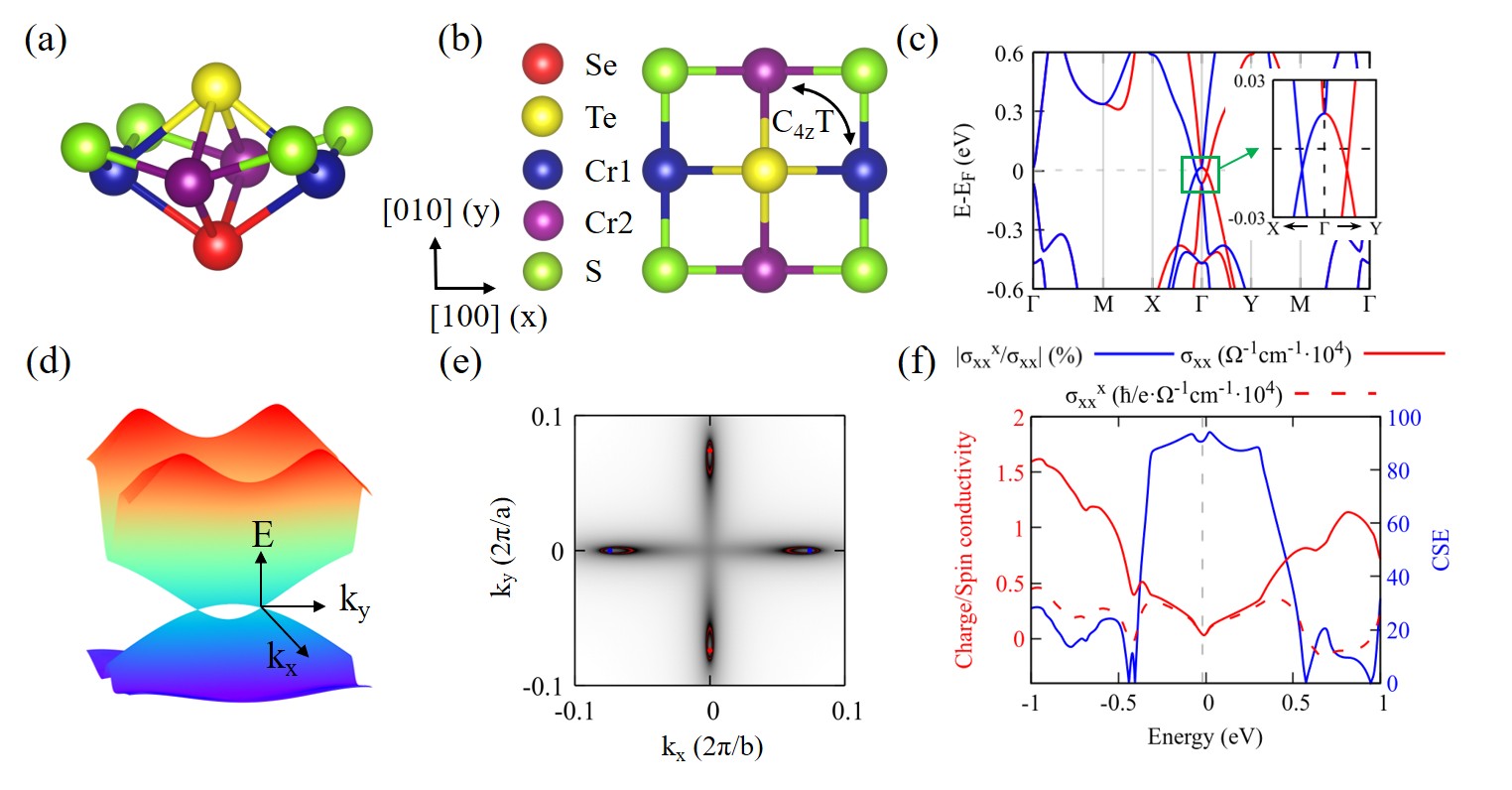}
\caption{\label{fig:cr2stese} Crystal structure, electronic structure, and spin transport properties of Cr$_2$SeTeS. Crystal structure of Cr$_2$SeTeS viewed from (a) the side view and (b) the top view. Cr1 and Cr2 label Cr atoms with oppositely oriented magnetic moments. (c) Spin-resolved band structure without spin-orbit coupling (SOC), red and blue lines indicate different spin orientations. (d) 3D band structure around the Dirac point, showing significant tilt of the Dirac cone. (e) Fermi surfaces at the charge neutrality point. Red and blue points indicate the Dirac points formed by opposite-spin bands. (f) Fermi-level-dependent charge conductivity, spin conductivity, and CSE of Cr$_2$SeTeS. The gray dashed line denotes the Dirac-point energy.}
\end{figure*}

Guided by the model analysis, we first examine the widely studied $d$-wave altermagnet Cr$_2$O$_2$, which has been shown to naturally host anisotropic Dirac cones and exhibit high mobility due to its steep linear dispersion \cite{guo2023qah,chen2023cro,xu2026monolayer}. Such high mobility is favorable for ultrafast channel transport, which can support short-pulse writing with a reduced driving current in device operation \cite{li2019materials,zhu2020energy}. The remaining question is whether this Dirac-cone geometry can generate both a sizable charge current and CSE.

Figures~\ref{fig:cr2o2}(a) and \ref{fig:cr2o2}(b) show the crystal structure of Cr$_2$O$_2$, where the two inequivalent Cr atoms in the primitive cell carry opposite spin moments. The spin-resolved band structure exhibits clean band crossings that lie nearly at the Fermi level [Fig.~\ref{fig:cr2o2}(c)]. Around the Dirac point, the Fermi surfaces at $E_F-0.1$ eV and $E_F+0.1$ eV are elliptical [Figs.~\ref{fig:cr2o2}(e) and \ref{fig:cr2o2}(f)]. The center of the ellipse is only weakly displaced from the Dirac point, indicating a small tilt. Therefore, Cr$_2$O$_2$ provides a material realization of the weakly tilted, anisotropic Dirac-cone regime in Fig.~\ref{fig:model_am}(b).

This correspondence is directly reflected in the transport response. As the Fermi level moves away from the Dirac point, both the charge conductivity and the spin conductivity increase, while the CSE remains nearly constant near the Dirac point [Fig.~\ref{fig:cr2o2}(g)]. The calculated CSE is 72\%, comparable to the 78\% value reported for the $d$-wave altermagnet KV$_2$Se$_2$O \cite{lai2025dwave}. Upon slight hole doping ($E-E_F=-98$ meV) and electron doping ($E-E_F=96$ meV), the charge conductivity of Cr$_2$O$_2$ is markedly enhanced while the CSE remains almost unchanged, driving the $T$-odd spin conductivity to the order of $10^3(\hbar/e)(\mathrm{S/cm})$. This value exceeds the spin Hall conductivity of 99.2\% of materials in the high-throughput spin Hall survey \cite{zhang2019spinorbitronic}. With the Fermi level shifted further away from the Dirac point, the charge and spin conductivities can be further enhanced nearly linearly, whereas the CSE remains almost unchanged. Thus, from the viewpoint of spin-source performance, Cr$_2$O$_2$ combines high charge conductivity with high CSE.

We further analyze how the spin-current tensor changes when the in-plane electric field is rotated, which is important for selecting an experimental writing geometry. The rotated tensor is obtained by \cite{seemann2015symmetry}
\begin{equation}
    \sigma_{i'j'}^{k'}=\sum_{lmn}R_{i'l}R_{j'm}R_{k'n}\sigma_{lm}^{n},
\end{equation}
where $R$ is the Euler rotation matrix. Unlike previously studied $d$-wave altermagnets characterized by easy-axis magnetocrystalline anisotropy \cite{lai2025dwave,cheng2026realistic}, Cr$_2$O$_2$ exhibits easy-plane anisotropy (Table S2 in Supplemental Material ~\cite{supplemental}). As shown in Figs.~\ref{fig:cr2o2}(h) and \ref{fig:cr2o2}(i), when the N\'eel vector is aligned along the crystallographic $a$ direction, the longitudinal CSE is maximized when the current flows along $a$, while a transverse component can be selected by rotating the current direction. In a simple device geometry, applying the current along $a$ directly accesses the maximum longitudinal CSE. 

The model further predicts that stronger anisotropy combined with finite tilt should push the CSE closer to the theoretical limit. To test this design rule, we propose the Janus monolayer Cr$_2$SeTeS, which is derived from layered altermagnetic V$_2$(Se,Te)$_2$O by replacing the transition-metal and chalcogen sublattices \cite{xu2026monolayer}. As shown in Figs.~\ref{fig:cr2stese}(a) and \ref{fig:cr2stese}(b), the Janus structure breaks the mirror symmetry between the two surfaces while preserving the compensated altermagnetic order.

Cr$_2$SeTeS realizes the desired stronger Dirac-cone deformation. The spin-resolved band structure shows Dirac points formed by opposite-spin bands along $X$--$\Gamma$ and $\Gamma$--$Y$ below the Fermi level [Fig.~\ref{fig:cr2stese}(c)]. The three-dimensional dispersion around the Dirac point is clearly tilted, and the slope along the $+k_y$ direction is much larger than that along the opposite direction [Fig.~\ref{fig:cr2stese}(d)]. Consistently, the Fermi surfaces in Fig.~\ref{fig:cr2stese}(e) show that the Dirac point is far from the center of the ellipse. The ellipticity is also stronger than that in Cr$_2$O$_2$, indicating a larger Dirac-cone anisotropy.

\begin{table}[t]
\caption{CSE of typical materials. NM, NC-AFM, AFM, and AM denote nonmagnetic, noncollinear antiferromagnetic, antiferromagnetic, and altermagnetic, respectively. SOC-SHC and SS-SC denote the spin Hall current induced by spin-orbit coupling and spin current induced by spin splitting, respectively.}
\label{tab:spin_conductivity_cse}
\begin{ruledtabular}
\begin{tabular}{lccc}
Material & Magnetism & Type & CSE (\%) \\
\hline
Pt \cite{liu2011stfmr} & NM & SOC-SHC  & 8 \\
$\beta$-W \cite{pai2012tungsten} & NM & SOC-SHC & 40 \\
Mn$_3$Sn \cite{zelezny2017spinpolarized} & NC-AFM &SS-SC & 15 \\
$\beta$-Fe$_2$PO$_5$ \cite{wang2026xtype} & X-type AFM & SS-SC & 80 \\
RuO$_2$ \cite{gonzalez2021efficient} &AM& SS-SC  & 27 \\
V$_2$Te$_2$O \cite{cui2023giant} & AM & SS-SC  & 32 \\
KV$_2$Se$_2$O \cite{lai2025dwave} & AM & SS-SC  & 78 \\
Cr$_2$O$_2^{*}$ & AM & SS-SC  & 72 \\
Cr$_2$SeTeS$^*$ &  AM &SS-SC  & 92 \\
\end{tabular}
\end{ruledtabular}
\end{table}

These band-structure features lead to a markedly enhanced CSE. As shown in Fig.~\ref{fig:cr2stese}(f), the charge and spin conductivity increase monotonically when the Fermi level moves away from the Dirac point, while the CSE first increases and then decreases, in good agreement with the tilted-cone behavior in Fig.~\ref{fig:model_am}(f). The CSE of 92\% at the charge neutrality point in Cr$_2$SeTeS exceeds those of heavy metals, noncollinear antiferromagnets, X-type antiferromagnets, and other altermagnets, as summarized in Table I. The largest value originates from the cooperative effect of strong anisotropy and substantial type-I tilt of the Dirac cone. With hole doping ($E-E_F=-53$ meV) and electron doping ($E-E_F=26$ meV), the charge conductivity is strongly enhanced. At the same time, the CSE increases to 94\%, together giving rise to a $T$-odd spin conductivity above $10^3(\hbar/e)(\mathrm{S/cm})$. Therefore, Cr$_2$SeTeS simultaneously exhibits high charge conductivity and a CSE approaching the quantum limit.

\textit{Conclusion}---In summary, we have proposed Dirac-cone engineering as a route to combine high charge conductivity with high CSE in 2D $d$-wave altermagnets. Model Hamiltonian calculations show that Dirac-cone anisotropy is the primary factor for enhancing CSE, while type-I tilt provides an additional tuning knob. Guided by this principle, we first examined Cr$_2$O$_2$, where a moderately anisotropic and weakly tilted Dirac cone gives rise to a CSE of about 72\%, close to the highest value reported so far. More importantly, we demonstrate that stronger anisotropy together with a sizable cone tilt in Cr$_2$SeTeS can further boost the CSE to 92\%, approaching the theoretical upper limit for nonrelativistic $T$-odd charge-to-spin conversion. In both materials, the intrinsically high carrier mobility associated with the Dirac cones allows slight carrier doping or electrostatic gating to provide a practical means to enhance the charge conductivity by shifting the Fermi level relative to the Dirac point, while maintaining a high CSE, thereby leading to a high spin conductivity. These findings establish the incorporation of Dirac cones into $d$-wave altermagnets as an effective strategy for reducing the writing current and energy dissipation in spintronic devices.

\textit{Acknowledgments}---We thank Ding-Fu Shao for useful discussions. This work was supported by the National Natural Science Foundation of China (Grants No. 52271016, No. 52188101 and No. 12504287) and Liaoning Provincial Doctoral Research Startup Fund Project (Grant No. 2025BS0176).


%

\end{document}